\documentclass[sigconf]{acmart}
\pdfoutput=1
\usepackage{booktabs} % For formal tables
\usepackage{tikz}
\usepackage{subfigure}
\usepackage{pgfplots}
\setcopyright{rightsretained}
\usetikzlibrary{patterns}

\definecolor{green1}{RGB}{204,203,102}
\definecolor{blue1}{RGB}{102,204,182}
\definecolor{red1}{RGB}{102,163,204}
\definecolor{purple1}{RGB}{102,33,102}
\definecolor{green2}{RGB}{255,153,102}
\definecolor{blue2}{RGB}{51,204,204}
\definecolor{red2}{RGB}{143,153,204}

\definecolor{tuatara}{RGB}{67, 67, 67}
\definecolor{aluminum}{RGB}{153,153,153}
\definecolor{silver}{RGB}{191,191,191}
\definecolor{platinum}{RGB}{228,227,228}
\definecolor{mercury}{RGB}{240,240,240}
\definecolor{gallery}{RGB}{250,250,250}
\definecolor{free_speech_aquamarine}{RGB}{0, 156, 114}
\definecolor{sun_shade}{RGB}{255, 144, 68}
\definecolor{fern}{RGB}{101,197,117}
\definecolor{french_blue}{RGB}{0, 112, 182}
\definecolor{sushi}{RGB}{117, 168, 47}
\definecolor{shakespeare}{RGB}{35, 184, 223}
\definecolor{egg_shell}{RGB}{238, 234, 215}
\definecolor{carnation}{RGB}{245, 80, 86}
\definecolor{flamingo}{RGB}{237, 88, 85}
\definecolor{jet_stream}{RGB}{188, 214, 210}
\definecolor{jelly_bean}{RGB}{45, 126, 150}
\definecolor{tree_poppy}{RGB}{246, 154, 27}
\definecolor{c1}{RGB}{255,97,56}
\definecolor{c2}{RGB}{255,255,157}
\definecolor{c3}{RGB}{190,235,159}
\definecolor{c4}{RGB}{121,189,143}
\definecolor{c5}{RGB}{0,163,136}

\author{Liang Pang, Yanyan Lan, Jiafeng Guo, Jun Xu, Xueqi Cheng}
\affiliation{%
  \institution{CAS Key Lab of Network Data Science and Technology, Institute of Computing Technology, \\ Chinese Academy of Sciences}
  \city{Beijing, China}
}
\email{pangliang@software.ict.ac.cn, {lanyanyan, guojiafeng, junxu, cxq}@ict.ac.cn}

\acmDOI{10.475/123_4}

\acmISBN{123-4567-24-567/08/06}

\copyrightyear{2017}
\acmYear{2017}
\setcopyright{rightsretained}
\acmConference{SIGIR 2017 Workshop on Neural Information Retrieval (Neu-IR'17)}{}{August 7--11, 2017, Shinjuku, Tokyo, Japan} 
\acmPrice{}
\begin{document}

\title{A Deep Investigation of Deep IR Models} 
%Explore the performance between representation based and interaction based matching models

\begin{abstract}
The effective of information retrieval (IR) systems have become more important than ever.
Deep IR models have gained increasing attention for its ability to automatically learning features from raw text; thus, many deep IR models have been proposed recently. However, the learning process of these deep IR models resemble a black box. Therefore, it is necessary to identify the difference between automatically learned features by deep IR models and hand-crafted features used in traditional learning to rank approaches. Furthermore, it is valuable to investigate the differences between these deep IR models.
This paper aims to conduct a deep investigation on deep IR models.  Specifically, we conduct an extensive empirical study on two different datasets, including Robust and LETOR4.0. 
We first compared the automatically learned features and hand-crafted features on the respects of query term coverage, document length, embeddings and robustness. It reveals a number of disadvantages compared with hand-crafted features. Therefore, we establish guidelines for improving existing deep IR models.
Furthermore, we compare two different categories of deep IR models, i.e. representation-focused models and interaction-focused models. It is shown that two types of deep IR models focus on different categories of words, including topic-related words and query-related words. 
\end{abstract}



% We no longer use \terms command


\maketitle

\section{Introduction}\label{sec:introduction}
Relevance ranking is the core problem in information retrieval (IR) system, which is to determine the relevance score for a document with respect to a particular query. Traditional approaches to tackle this problem include heuristic retrieval models and learning to rank approach. Heuristic retrieval models, such as TF-IDF~\cite{salton1986introduction} and BM25~\cite{robertson1994some}, propose to incorporate human knowledge on relevance into the design of ranking function. Modern learning to rank approach currently turns to apply machine learning techniques to the ranking function, which combines different kinds of human knowledge (relevance features such as BM25 and PageRank) and therefore has achieved great improvements on the ranking performances~\cite{liu2009learning}. However, a successful learning to rank algorithm usually relies on effective hand-crafted features for the learning process. The feature engineering work is usually time consuming, incomplete and over-specified, which largely hinder the further development of this approach~\cite{DRMM}.

Deep IR models have gained increasing attention for its ability to automatically learning features from raw text of query and document. Therefore, many deep IR models have been proposed to solve relevance ranking problem only considering the query and document textual data. As \cite{DRMM} has mentioned, deep IR models can be categorized into two branches, namely representation-focused models and interaction-focused models, depending on the different structures. However, deep IR models are the end-to-end system, the learning process of the deep IR models are still resemble a black box. Thus, it is curious to us that the different between automatically learned features by deep IR models and hand-crafted features used in traditional learning to rank approaches, and the differences between these deep IR models.

In this paper, we conduct a deep investigation on deep IR models under two aspects. 
Firstly, we compare the automatically learned features with hand-crafted features. Hand-crafted features, including TF-IDF, BM25 and other traditional retrieval models, are combined human knowledge and proven to follow some heuristic retrieval constrains~\cite{fang2004formal}. Incorporate with these constrains, we check the properties of deep IR models, for example query term coverage, document length and embedding affect the performance of deep IR models. Additionally, We pick out bad cases from test dataset to conduct error analysis. For each group of bad cases, we establish guidelines for improving existing deep IR models. We also show that the robustness of automatically features are stronger than hand-crafted features. 
Secondly, we compare the differences between two kinds of deep IR models. By visualizing the pooling words of these deep IR models, it is shown that representation-focused models focus on topic-related words and interaction-focused models focus on query-related words. A synthetic experiment further prove these properties. 

The rest of the paper is organized as follows. In Section~\ref{sec:deep-ir-model}, we introduce two types of existing deep IR models. In Section~\ref{sec:dataset}, we introduce the datasets for performance evaluation. Section~\ref{sec:part1} compares the differences between the automatically learned features and the hand-crafted features and  Section~\ref{sec:part2} compares the differences between two types of deep IR models. Section~\ref{sec:conclusion} concludes the paper.

\section{Existing Deep IR Model}\label{sec:deep-ir-model}
The core problem of information retrieval is to determine the relevance score for a document with respect to a particular query, which can be formalized as the follows as indicated in \cite{ARC-II} and \cite{MatchPyramid}. Given a query $Q=\{q_1,\cdots,q_m\}$ and a document $D=\{w_1,\cdots,w_n\}$, where $q_i$ and $w_j$ stand for the $i$-th and $j$-th words in the query and document respectively, the degree of relevance is usually measured as a score produced by a scoring function based on the representations of the query and document:

	\begin{equation}
		match(Q,D)=F(\Phi(Q),\Phi(D)),
	\end{equation}
where $\Phi$ is a function to map query/document to a vector, and $F$ is a scoring function for modeling the interactions between them. 

Deep IR models propose to automatically learn relevance features from raw text data, i.e.~$Q$ and $D$. Considering different structures, existing deep models can be categorized into two kinds: representation-focused models and interaction-focused models. The representation-focused models propose to focus on the learning parameters of function $\Phi$, while interaction-focused models put more efforts on learning parameters of function $F$.

\subsection{Representation-Focused Models}
\begin{figure}
	\centering
	\includegraphics[width=0.40\textwidth]{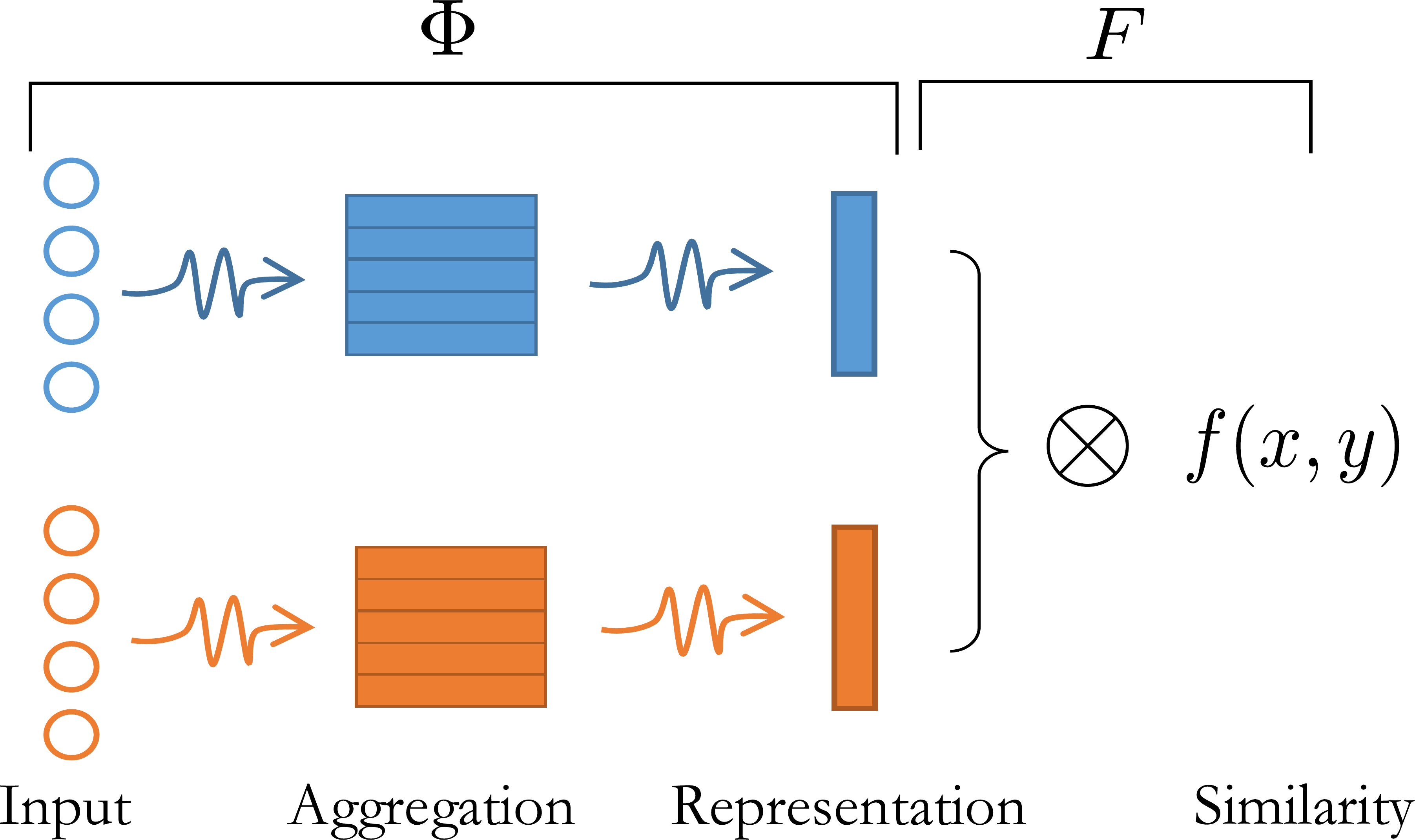}	
	\caption{Representation-Focused Models}
	\label{Fig.siamese}
\end{figure}

Representation-focused models try to build a good representation for query/document with a deep neural network, and then conduct matching between two abstract representation vector. In this approach, $\Phi$ is a relatively complex representation mapping function from text to vector, while $F$ is a relatively simple matching function. For example, in DSSM~\cite{DSSM}, $\Phi$ is a feed forward neural network with letter trigram representation as the input, while $F$ is the cosine similarity function. In CDSSM~\cite{CDSSM}, $\Phi$ is a convolutional neural network (CNN) with letter trigram representation as the input, while $F$ is the cosine similarity function. In ARC-I~\cite{ARC-II}, $\Phi$ is a CNN with word embeddings as the input, while $F$ is a multi-layer perceptron (MLP). Without loss of generality, all the model architectures of representation-focused models can be viewed as a Siamese (symmetric) architecture over the text inputs, as shown in Figure~\ref{Fig.siamese}. 

\subsection{Interaction-Focused Models}
Interaction-focused models first build the local interactions between query and document, based on basic representations, and then use deep neural networks to learn the complex interaction patterns for relevance. In this approach, $\Phi$ is usually a simple mapping function to map query and document to a sequence of words or word vectors, while $F$ is a complex deep model with many learnable parameters. Typically, function $F$ can be represented as the compound function of $H$, $G$ and $M$, i.e.~$F=H\circ G\circ M$, and the scoring function can be written as the following form:
	\begin{equation}
		match(Q,D)=H\circ G\circ M(\Phi(Q),\Phi(D)),
	\end{equation}
where $M$ is a function used to obtain the local interactions between the representation of $Q$ and $D$, $H$ is a deep neural network to obtain the abstract interaction patterns, and $H$ is an aggregation function to obtain the relevance score based on the interaction patterns. 

\begin{figure}
	\centering
	\includegraphics[width=0.45\textwidth]{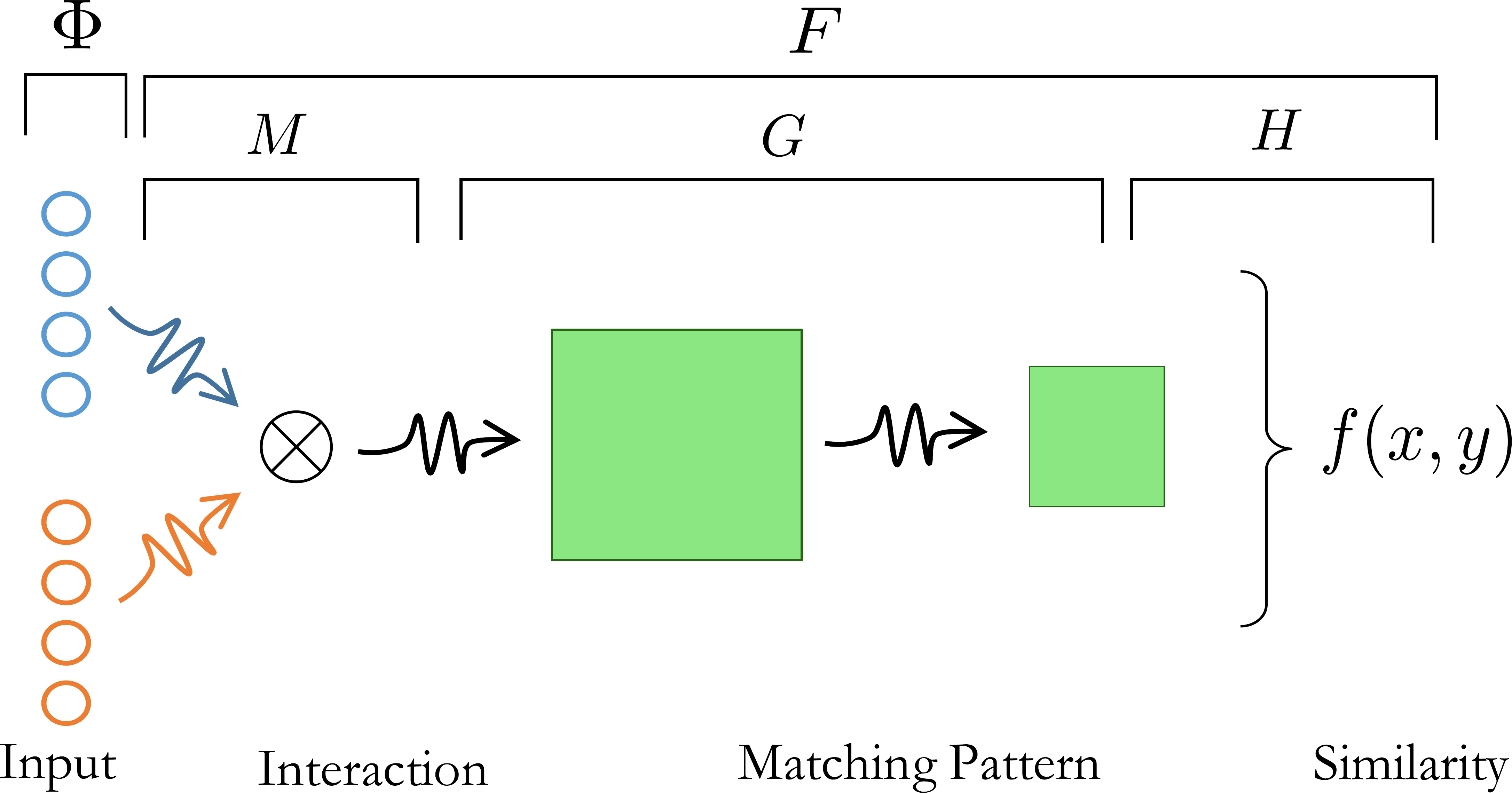}	
\caption{Interaction-Foused Models}
	\label{Fig.match_ptn}
\end{figure}

For example, in DeepMatch~\cite{DeepMatch}, $\Phi$ is an identical function which maintains the representation of query/document as a sequences of words, $M$ is a basic interaction function to output parallel texts, i.e.~the set of interacting pairs of words from $Q$ and $D$, $G$ is a feed forward neural network constructed by a topic model over the parallel texts, and $H$ is a logistic regression unit to summarize the decision to obtain the final relevance score. In ARC-II~\cite{ARC-II}, $\Phi$ is a mapping function to map query/document to a sequence of word embeddings, $M$ is the 1-D convolution operation over each patch of words from $Q$ and $D$, $G$ is a CNN, and $H$ is an MLP.  In MatchPyramid~\cite{MatchPyramid}, $\Phi$ maps query/document to a sequence of word vectors, $M$ is the similarity function between each word pair from $Q$ and $D$ to output a word-level interaction matrix, $G$ is a CNN, and $H$ is an MLP. In Match-SRNN, $\Phi$ is the same as that in MatchPyramid, $M$ is a tensor operation to incorporate high dimensional word level interactions, $G$ is a 2D-GRU, and $H$ is an MLP. Without loss of generality, all the interaction focused models can be viewed as a deep architecture over the local interaction matrix, as shown in Figure~\ref{Fig.match_ptn}.

\section{Dataset} \label{sec:dataset}
In this section, we introduce two datasets used for model analysis, namely Robust Dataset, LETOR 4.0 Dataset. The statistics of these datasets are shown in Table~\ref{Table.Dataset}. We can see that these datasets represent different sizes and genres of heterogeneous text collections. 

\begin{table}[!htbp]
    \centering
    \caption{Statistics on Robust and LETOR 4.0 .}
	\label{Table.Dataset}

	\begin{tabular}{l r r r}
		\hline
		Dataset & \#Query  & \#Doc & \#Relevance\\
		\hline
		Robust & 250 & 12240 & 12881 \\
		LETOR 4.0 & 1501 & 57899 & 61480 \\
		\hline
	\end{tabular}
\end{table}

\subsection{Robust Dataset}
Robust data is a small news dataset. Its topics are collected from TREC Robust Track 2004. We made use of the title of each TREC topic in our experiments. The retrieval experiments on this dataset are implemented using the Galago Search Engine\footnote{\url{http://www.lemurproject.org/galago.php}}. During indexing and retrieval, both documents and query words are white-space tokenized, lower-cased, and stemmed using the Krovetz stemmer. Stopword removal is performed on query words during retrieval using the INQUERY stop list.

\subsection{LETOR 4.0 Dataset}
LETOR4.0 dataset~\cite{qin2010letor} is a benchmark data for evaluating learning to rank methods. This dataset is sampled from the .GOV2 corpus using the TREC 2007 Million Query track queries. This dataset contains two subsets, i.e.~MQ2007 and MQ2008. In this paper, we use MQ2007 for evaluation because it is much larger than MQ2008. MQ2007 contains 1692 queries and 65,323 documents, which is much larger than Robust. Each query and document pair in this dataset is represented as a vector using 46 different features. The separation of training, validation and testing set are set to default. The reason to choose LETOR4.0 beyond Robust lie in that: 1) the LETOR4.0 data is relatively large (especially for the query number), therefore it is more appropriate for training a deep learning model; 2) the features have already been extracted, therefore it is convenient to conduct comparisons with learning to rank baselines.

\section{Comparisons with Hand-crafted Features} \label{sec:part1}
Learning to rank approaches with hand-crafted features have achieved a great success in information retrieval. In these hand-crafted features, BM25 and language model are the strong baselines in information retrieval. It is mainly because these traditional models satisfied several heuristic retrieval constrains proposed by Fang et al.~\cite{fang2004formal}. These constrains reveal importance properties in information retrieval. They are 
1) Term Frequency Constraints (TFC1 / TCF2); 2) Term Discrimination Constraint (TDC); 3) Length Normalization Constraints (LNC1 / LNC2); 4) TF-LENGTH Constraint (TF-LNC). 
Empirical results show that when a constraint is not satisfied, it often indicates non-optimality of the method. 

In this section, our aim is to find out the differences between deep IR models and hand-crafted features. Firstly, bad cases are categorized to identify the weakness of deep IR models. Then we make use of heuristic retrieval constrains to explain the disadvantages of the deep IR models, such as query term coverage problem, document length problem and embedding semantic abuse problem. Lastly, we point out that the advantage of the deep IR models is the robustness of the automatically learnt features. 

\subsection{Performance Comparison}
Before we conduct error analysis, we first overview the performance of two kinds of deep IR models and two strong hand-crafted features, namely BM25 and language model, shown in Table~\ref{Table.Experiments2}. 

The experimental result shows that deep IR models perform worse than the hand-crafted features, especially compared with BM25. Furthermore, we find that the performance gap between deep IR models and hand-crafted features is large in Robust dataset compared with LETOR 4.0 dataset. The main reason lays to the small size of the dataset, which go against the data-driven mechanism in deep learning. We reduce the performance gap by increasing the size of the dataset in LETOR 4.0, and we believe that the larger dataset will result in better performance, even better than the hand-crafted features.

Apart from the affection of dataset size, other heuristic problems are found by conducting error analysis. We will demonstrate these heuristic problem in the next section. 

\begin{table}
    \centering
    \caption{Performance comparison of deep IR models and hand-crafted features on Robust, LETOR 4.0.}
	\label{Table.Experiments2}

	\begin{tabular}{l l l l l l}
		\multicolumn{6}{c}{Robust} \\
		\hline
		Model & NDCG@1 & NDCG@10 & P@1 & P@10 & MAP\\
		\hline
		\hline
		\textsc{BM25-Title} & 0.563 & 0.445 & 0.563 & 0.402 & 0.255 \\
		\textsc{LM.JM-Title} & 0.560 & 0.443 & 0.560 & 0.400 & 0.253\\
		\hline
		\textsc{Arc-I} & 0.124 & 0.138 & 0.124 & 0.132 & 0.050 \\ % 300
		\textsc{MatchPyramid} & 0.364 & 0.242 & 0.364 & 0.240 & 0.164 \\ % 10000
		\hline	
		\multicolumn{6}{c}{LETOR 4.0} \\
		\hline
		Model & NDCG@1 & NDCG@10 & P@1 & P@10 & MAP\\
		\hline
		\hline
		\textsc{BM25-Title} & 0.358 & 0.414 & 0.427 & 0.366 & 0.450 \\
		\textsc{LM.JM-Title} & 0.300 & 0.374 & 0.359 & 0.329 & 0.421\\
		\hline
		\textsc{Arc-I} & 0.310 & 0.386 & 0.376 & 0.364 & 0.417 \\ % 300
		\textsc{MatchPyramid} & 0.362 & 0.409 & 0.428 & 0.371 & 0.434 \\ % 10000
		\hline	
	\end{tabular}
\end{table}

\subsection{Error Analysis}

\subsubsection{Query term coverage problem}
With analyzing number of bad cases of deep IR model results, we find that in the most of the cases, deep IR models are hard to satisfy the Term Discrimination Constraint (TDC)~\cite{fang2004formal}. TDC ensures that given a fixed number of occurrences of query terms, we favor a document that has more occurrences of discriminative terms. 
However, for both Arc-I and MatchPyramid, we never intentionally designed a network structure to distinguish two same query terms matching from two discriminative query terms matching. Although, it is possible for these two models to learn from data, the limitation is the size of the dataset, which make it hard for deep IR models to learn TDC without any prior information.

We pick out one example, shown in Figure~\ref{Figure.QueryCoverage}, to demonstrate this issue. Here the query is ``tooth fairy museum''. In the left part of Figure~\ref{Figure.QueryCoverage}, the document only contains query terms ``museum'' and ``fairy'', where term ``fairy'' occurs more than 10 times. While in the right part of Figure~\ref{Figure.QueryCoverage}, the document contains all three query terms, but the total number of query terms is only 6 times. With the constrain of TDC, a good IR model prefer to rank right side document higher than left side. However, the deep IR models make the opposite decision.

\textbf{Suggestions}: In order to recover query term coverage problem we propose two suggestions for representation-focused model and interaction-focused model respectively.
1) For representation-focused models, an attention mechanism turns to be a useful strategy to distinguish different query terms. 
2) For interaction-focused models, pooling across each query term rows in the interaction matrix is helpful for considering different query terms individually.

\begin{figure*}
    \begin{tikzpicture}
        \node[text width=0.45\textwidth] at (0,0) {... library {\color{blue}museum} legend christmas {\color{green}fairy} legend christmas {\color{green}fairy} word {\color{green}fairy} come latin word fata meaning fate means {\color{green}fairy} cousin classical fates believe control fate destiny human race hope {\color{green}fairy} associate christmas good ... good men germany christmas {\color{green}fairy} legend tell ... patter small feet coming hall room open door {\color{green}fairy} clad sparkle robe dance laugh singing splendid ... ernestine queen {\color{green}fairy} came return lost gold ring love sight count otto soon ask {\color{green}fairy} queen bride ... live happily years count otto {\color{green}fairy} wife decided hunt forest near castle count grew impatient ...};
        \node[text width=0.45\textwidth] at (0.5\textwidth,0) {... dr samuel harris national {\color{blue}museum} dentistry baltimore made possible generous support colgate palmolive children {\color{blue}museum} virginia select nine children science {\color{blue}museum} host branch bristle ... toothbrush ancient babylonia st century {\color{red}tooth} sticks made wild ... proper space inspire visitor make healthy snack choice computer driven essential toothbrush {\color{red}tooth} {\color{green}fairy} tech savvy visitor option create virtual toothbrush learn evolutionary time periods emphasize dimension timeline visitor take time innovate design handcrafte toothbrush bench view vintage dental product poster find ... };   
    \end{tikzpicture}
    \caption{Two documents related to query ``tooth fairy museum'', the right document cover one more query term than the left document, while the left document rank high in the deep IR models.}
    \label{Figure.QueryCoverage}
\end{figure*}

\subsubsection{Document length problem}
Another type of errors relates to the general preprocessing of document length limitation before feed it into the deep IR models. With the limitation of memory and time, documents tailor to a maximum length. It always make sense in paraphrase identification tasks and question answering tasks, which have the similar text length. But in information retrieval tasks, documents have variance length, ranging from 10 to 10,000 words. Directly cut off the exceeded text leads to information loss and violates the Term Frequency Constrain (TF1). TF1 claims that replacing one of the non-query term word to a query term increases the relevance score. Therefore for a cut-offed document, if the replacement occurs in the exceeded part, the relevance score keeps the same in deep IR models.

\begin{figure*}
  \centering
  \includegraphics[width=0.9\textwidth]{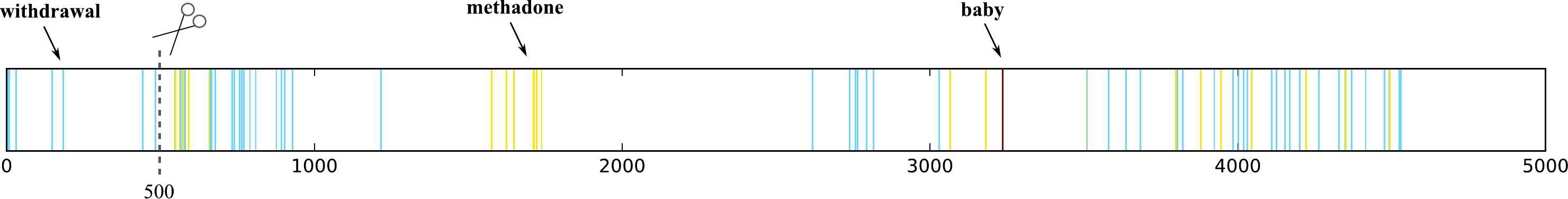}
  \caption{An example document shows that the exceed part of a long document has much query term matching information.}
  \label{Figure.DocLengthExp}
\end{figure*}

For example, a document contains 5000 words in Figure~\ref{Figure.DocLengthExp}, and the maximum length of our model set to 500 words. Thus we find that in the view of deep IR models, only few query terms ``withdrawal'' occur in the top 500 words of the document. While, as the figure shown, the query terms ``methadone'' and ``baby'' occur in the exceeded part of the document.

\begin{figure}
  \centering
  \includegraphics[width=0.45\textwidth]{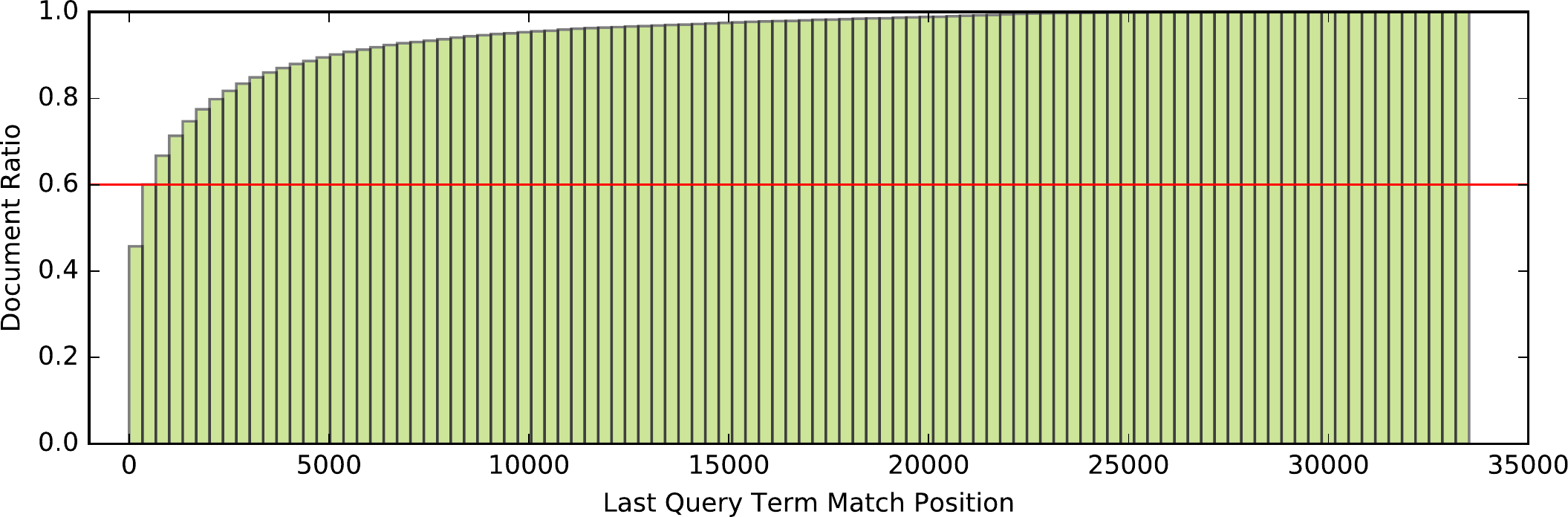}
  \caption{The distribution of last query term matching position.}
  \label{Figure.DocLengthDist}
\end{figure}

A more precision statistic of the last query term match position in a document is shown in Figure~\ref{Figure.DocLengthDist}. The red line represents the 500 words threshold, thus about 40\% of document loss the query term information because of the length limitation of the document. So this special preprocess affect 40\% of the documents, which reflecting on the worse model performance.

\textbf{Suggestions}: For this situation, inspired by passage retrieval approaches, document can be split into several short passages. Then deep IR models apply on each <query, passage> pairs. Finally, the relevance score is the aggregation of each <query, passage> local relevance scores.

\subsubsection{Embedding semantic abuse problem}
Different from using one hot word representation in the hand-craft features such as BM25 and language model, recently most deep IR models adopt pre-trained word embeddings as the word representation, except the letter-trigrams used in DSSM and CDSSM. The advantage of adopting word embedding as the word representation is to investigate semantic matching information into the model. However, on the flip side, semantic matching brings too much noise matching signals, which covers up the exact matching signals and dominates the final matching score. Similar to the heuristic constrains proposed in \cite{fang2004formal} which only consider exact matching signals, we append one constrain by considering semantic matching signals, called Term Semantic Frequency Constrain (TSFC). 

\textbf{TSFC:}  Let $q = \{w\}$ be a query with only one term $w$. Assume $|d1| = |d2|$ and $s(w,d1) = s(w,d2)$. If $c(w,d1) > c(w,d2)$, then $f(d1,q) > f(d2,q)$.

This constrain assumes that two documents have the same length, and the sum of the semantic matching signals (including exact matching signals) are equivalent, $s(w,d1) = s(w,d2)$. The larger number of the exact matching signals $c(w,d)$, the higher relevance score $f(d,q)$.

\begin{figure}
  \centering
  \includegraphics[width=0.48\textwidth]{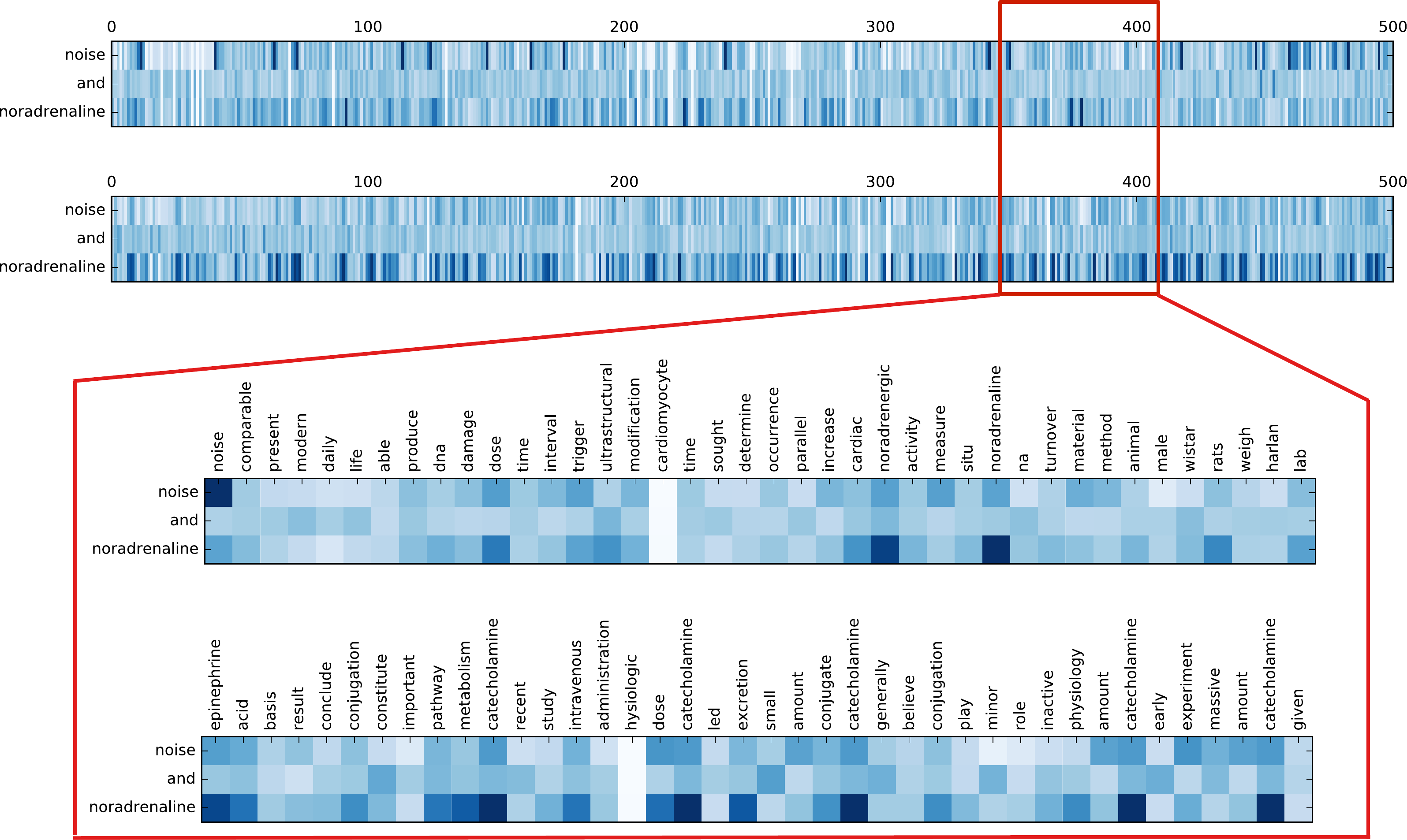}
  \caption{An example illustrates the embedding semantic matching problem. The first document contains exact matching signals, while the second document contains much high semantic matching signals.}
  \label{Figure.Embedding}
\end{figure}

Take an instance in LETOR 4.0 Dataset as an example shown in Figure~\ref{Figure.Embedding}. The given query term ``noradrenaline'' has a high similarity with so many medical related words, such as ``epinephrine'', ``catecholamine'' and ``metabolism''. Thus the sum of matching signals in the lower document is higher than that in the upper document, even the lower document dose not has any exact matching signal with query term ``noradrenaline''. As the experiment shows that deep IR models prefer the lower document, for the sake of the higher density matching signals.

\begin{figure}
		\centering
		\subfigure[Dot Product]{
			\label{Fig.sub.1}
			\includegraphics[width=0.31\textwidth]{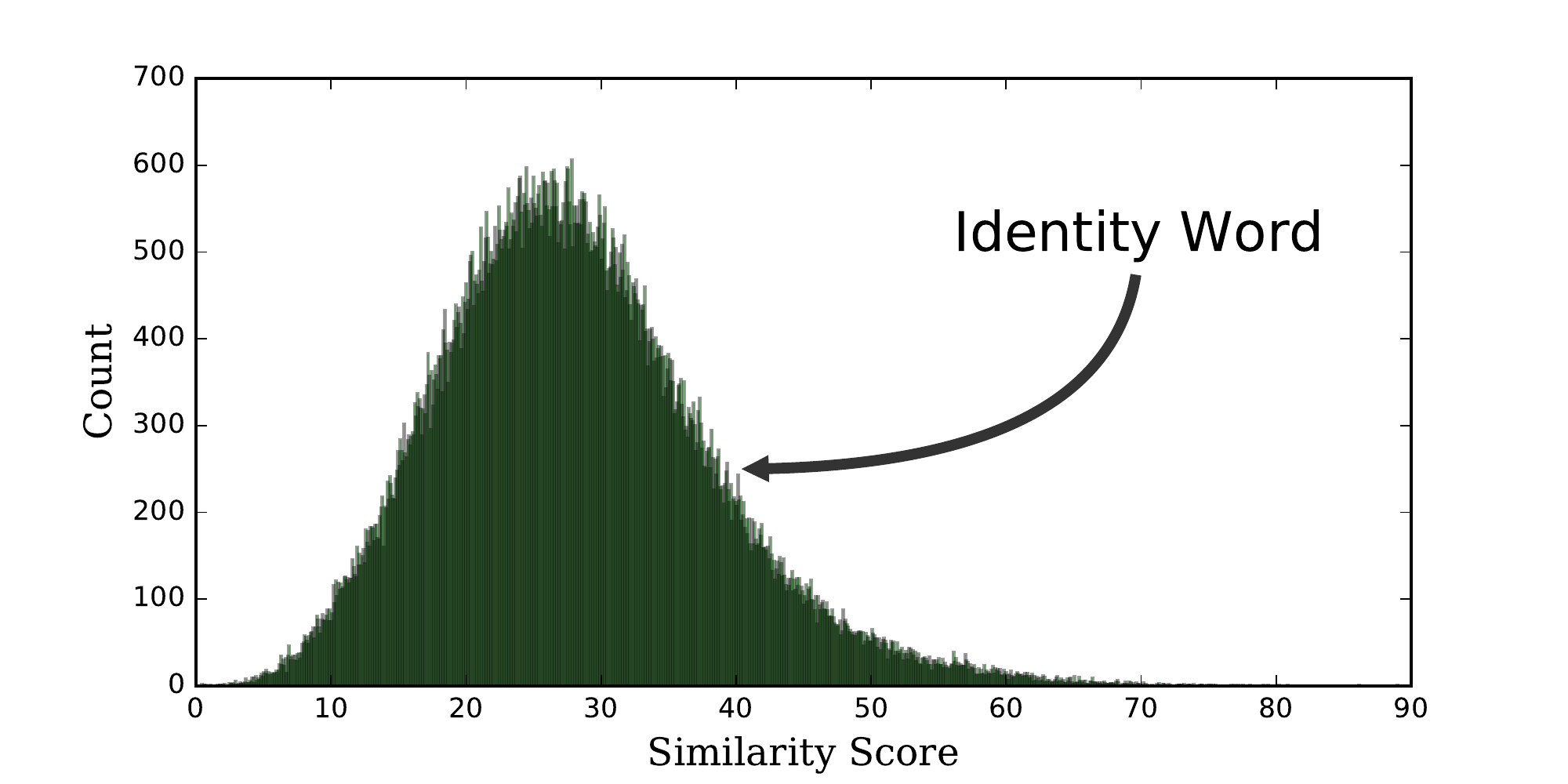}
			}
		\subfigure[Cosine]{
			\label{Fig.sub.2}
			\includegraphics[width=0.31\textwidth]{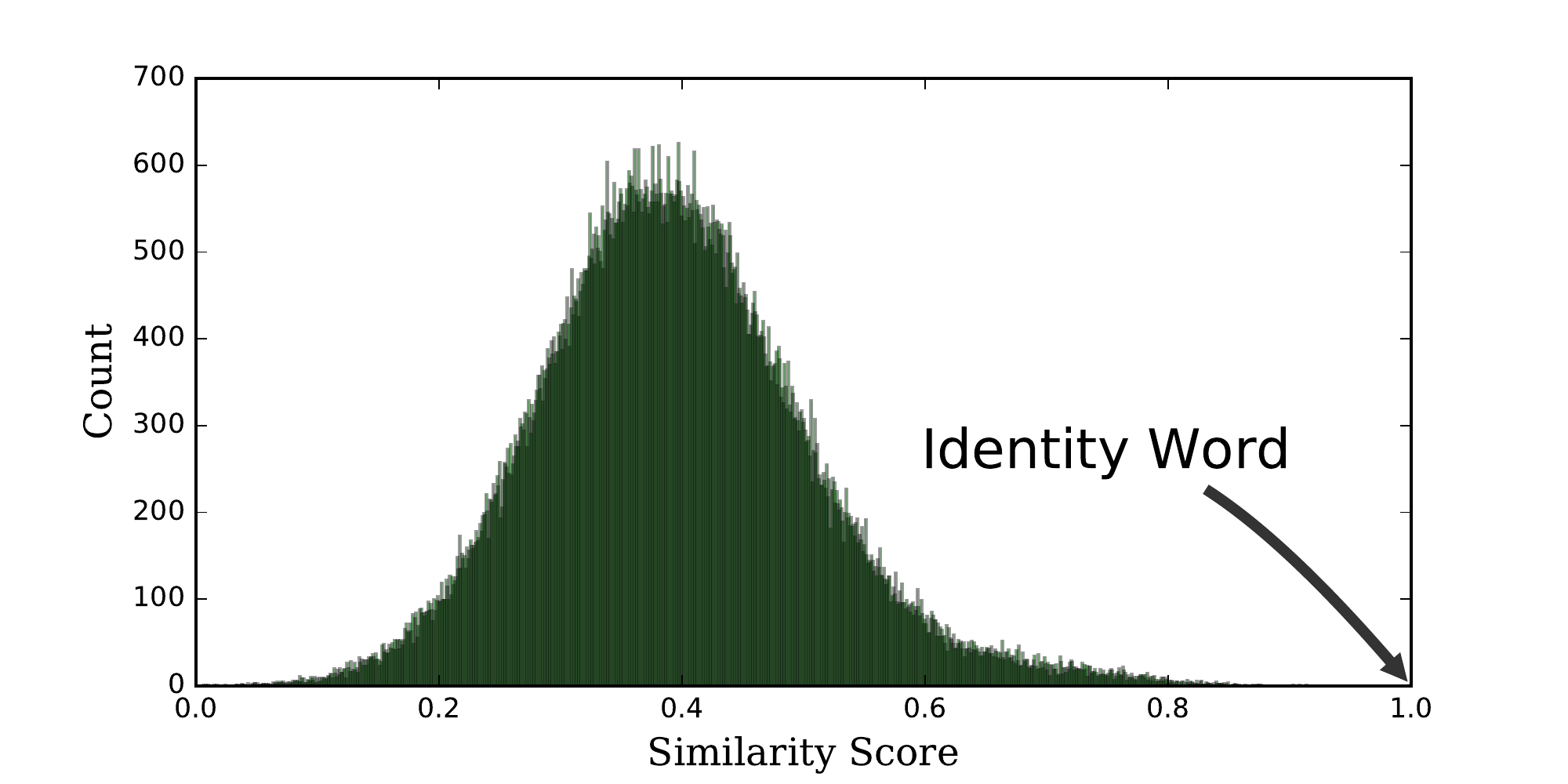}
 		}
 		\subfigure[Gaussian Kernel]{
			\label{Fig.sub.3}
			\includegraphics[width=0.31\textwidth]{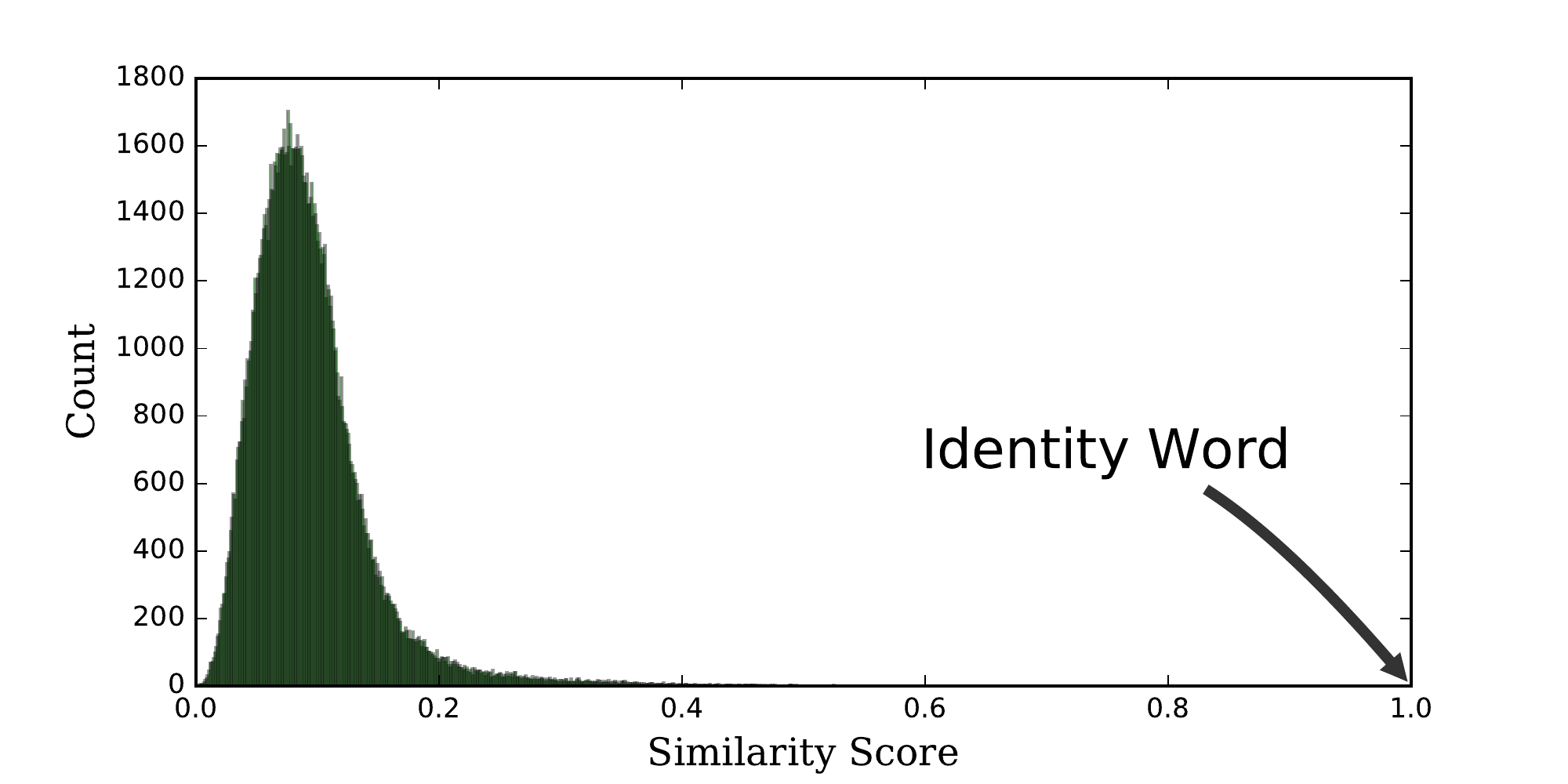}
 		}
 		\caption{Choose one word from the vocabulary and measure the similarity between other words, we draw the histogram of three type of similarity functions: dot product, cosine and gaussian kernel. The arrow point the similarity between two identity word (the word we choose).}
 		\label{Fig.word_dist}
	\end{figure}

\textbf{Suggestions}: As TSFC shown that we need enlarge the gap between semantic matching signals and exact matching signals. As the Figure~\ref{Fig.word_dist} shows that for interaction-focused models, we can define a proper similarity function between words, so that the exact matching signals are larger than all semantic matching signals, such as the similarity functions show in Figure~\ref{Fig.sub.2} and Figure~\ref{Fig.sub.3}.
	
\subsubsection{Feature Robustness}
The robustness of the features can be interpreted as that when some of the features are missing, how much it affects the model performance. The features we take into consideration are the 46 dimensional features provided in LETOR 4.0 dataset and the last layer 20 dimensional outputs in the Arc-I and MatchPyramid. 
Then we use a linear model to fit these 3 sets of features to the final relevance labels. In this way, the learnt weights reflect the importance of the features.

\begin{figure*}
  \centering
  \includegraphics[width=0.9\textwidth]{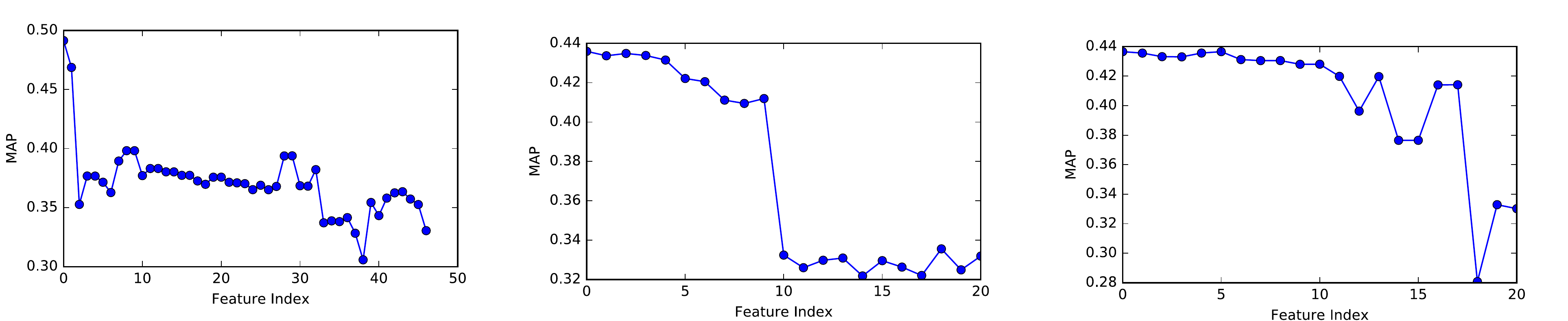}
  \caption{Left: 46 dimensional hand-crafted features; Middle: Arc-I last layer 20 dimensional features; Top: MatchPyramid last layer 20 dimensional features.}
  \label{Figure.Robustness}
\end{figure*}

In order to visualize the robustness different between hand-crafted features, Arc-I and MatchPyramid, we conduct experiments on the first fold of LETOR 4.0 dataset. After sufficient model training, we firstly sort features by its importance (their corresponding weight). Then remove features one-by-one following the order of the feature importance at each time. Finally, evaluate the performance of the rest of the features. The Figure~\ref{Figure.Robustness} illustrates the procedure of the feature removing. Figure~\ref{Figure.Robustness}(a) shows the result of hand-crafted features, as we can see that when we remove the first two features the performance drops a lot to about 0.36. Figure~\ref{Figure.Robustness}(b-c) shows the results of Arc-I and MatchPyramid, as we can see that even half of the feature have been removed, the model performance affects a little. That is to say, deep IR models automatically learnt features turn to be more robustness than the hand-crafted features.

\section{Comparisons Among Deep IR Models} \label{sec:part2}
In this section, we investigate the differences between representation-focused models and interaction-focused models based on information retrieval task. Deep IR models lay into these two categories, such as DSSM, CDSSM and Arc-I belong to representation-focused models, Arc-II and MatchPyramid belong to interaction-focused models. We choose two classical models from each category, Arc-I and MatchPyramid. In order to explorer their intrinsic differences, we conduct our analysis on Robust dataset, LETOR 4.0 dataset and a simulated dataset.

\subsection{Performance Comparison}
\begin{table}
    \centering
    \caption{Performance comparison of different deep IR models on Robust, LETOR 4.0.}
	\label{Table.Experiments1}

	\begin{tabular}{l l l l l l}
		\multicolumn{6}{c}{Robust} \\
		\hline
		Model & NDCG@1 & NDCG@10 & P@1 & P@10 & MAP\\
		\hline
		\hline
		\textsc{DSSM} & 0.122 & 0.137 & 0.122 & 0.135 & 0.048\\
		\textsc{CDSSM} & 0.118 & 0.134 & 0.118 & 0.130 & 0.042 \\ % 100
		\textsc{Arc-I} & 0.124 & 0.138 & 0.124 & 0.132 & 0.050 \\ % 300
		\hline
		\textsc{Arc-II} & 0.140 & 0.148 & 0.140 & 0.156 & 0.054\\ % 800
		\textsc{MatchPyramid} & 0.364 & 0.242 & 0.364 & 0.240 & 0.164 \\ % 10000
		\hline
		\multicolumn{6}{c}{LETOR 4.0} \\
		\hline
		Model & NDCG@1 & NDCG@10 & P@1 & P@10 & MAP\\
		\hline
		\hline
		\textsc{DSSM} & 0.290 & 0.371 & 0.345 & 0.352 & 0.409 \\ % 100
		\textsc{CDSSM} & 0.288 & 0.325 & 0.333 & 0.291 & 0.364\\
		\textsc{Arc-I} & 0.310 & 0.386 & 0.376 & 0.364 & 0.417 \\ % 300
		\hline
		\textsc{Arc-II} & 0.317 & 0.390 & 0.379 & 0.366 & 0.421\\ % 800
		\textsc{MatchPyramid} & 0.362 & 0.409 & 0.428 & 0.371 & 0.434 \\ % 10000
		\hline		
	\end{tabular}
\end{table}

The performance comparison results illustrate in Table.~\ref{Table.Experiments1}. 
After comparing different models on different datasets, we can conclude as follow:
1) interaction-focused models are performed better than representation-focused models;
2) the gap between representation-focused models and interaction-focused models on dataset Robust is larger than the gap on dataset LETOR 4.0.

It motives us to explore the differences between representation-focused models and interaction-focused models.

\subsection{Property Analysis}

% query 短， document 长，压缩成主题，做匹配
% 对于长度相当的句子，压缩主题容易，对于query，会很困难
\subsubsection{Text Representations}
Text representations is a major task for representation-focused models, since text representation is used to compress most valuable and distinguishable information into one vector. In paraphrase identification task, two sentences are symmetric and have similar length. 
However, in information retrieval, query and document are totally different objects. 
Query is abstract and almost every terms in it reflect a perspective of search intent. Thus, text representation for a query need to keep all the query terms information in one vector.
On contrast, document is elaborate and relevant document just follow two hypotheses in the literature~\cite{robertson1994some}. The Verbosity Hypothesis assumes that a long document is like a short document, covering a similar scope but with more words; while the Scope Hypothesis assumes a long document consists of a number of unrelated short documents concatenated together. Thus, text representation for a document only need to contain the important part of a document.

In order to analyze the information encoded in text representation, we visualize the words at the max pooling position in Arc-I model. 

\begin{figure}
  \centering
  \includegraphics[width=0.45\textwidth]{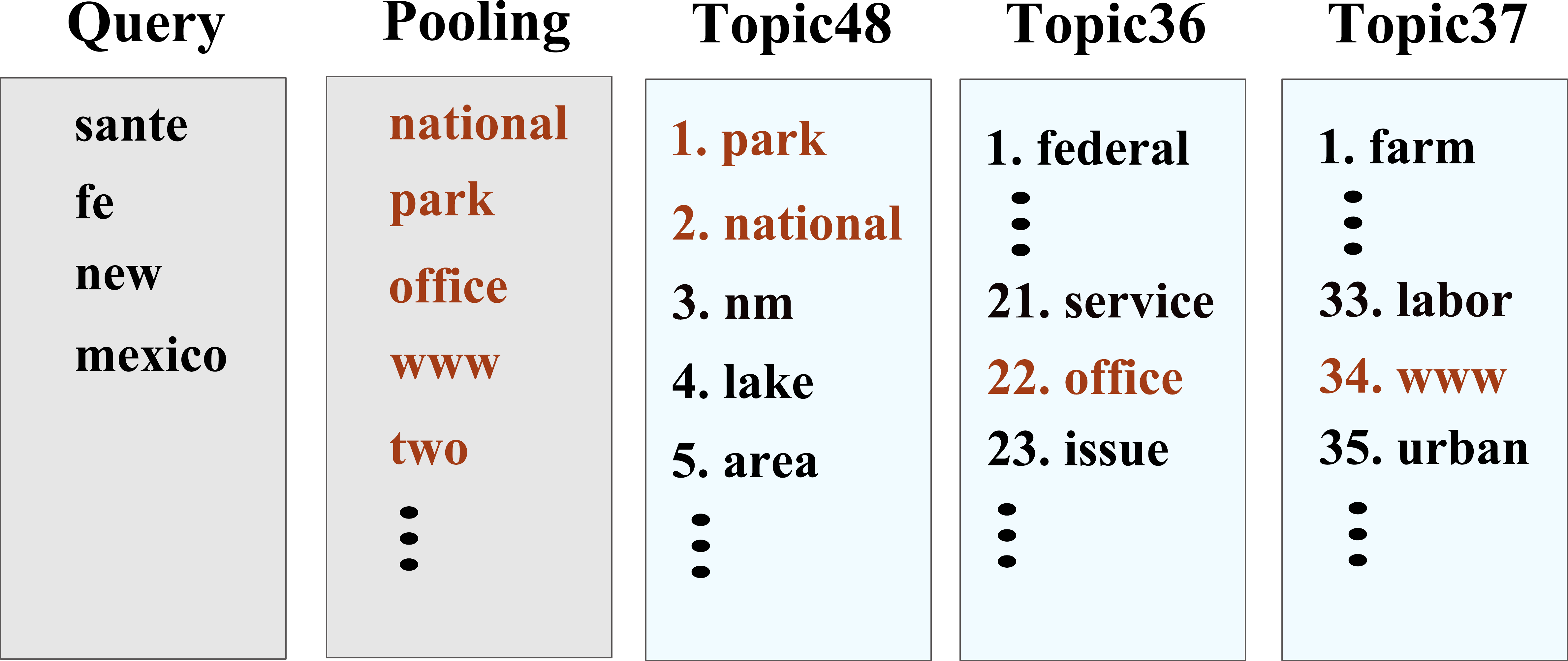}
  \caption{The Arc-I pooling words have relation to LDA topic words.}
  \label{fig.arc1-pooling-text}
\end{figure}

As an example, in Figure~\ref{fig.arc1-pooling-text}, the query terms are ``\textbf{sante fe new mexico}'', which are listed in the first text box. The words selected by max pooling layer based on convolution output feature maps, are listed in second text box, such as ``national'' ``part''. It is evident to see, all of the pooling words are not the words appear in query terms. The reason of the above observation is that the document and query text representations are extracted independently, thus the relation of these two representations are weak. For document, most informative words are extracted to composite the representation, which have the same purpose with the topic models. Thus we assume that the pooling words are related to the topic words in the topic models.

\begin{figure}
  \centering
  \includegraphics[width=0.46\textwidth]{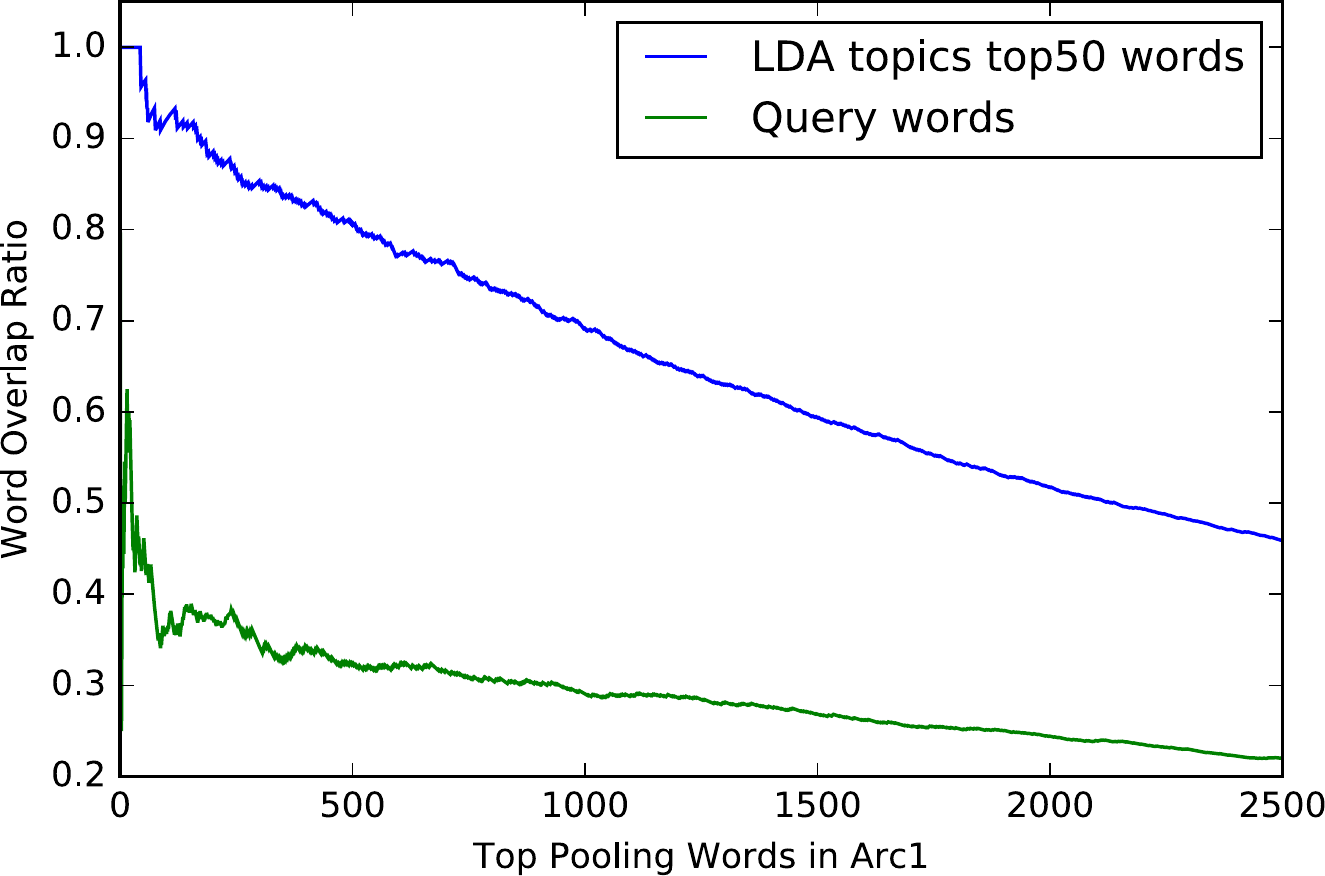}
  \caption{The overlap ratio of top pooling words in Arc-I with LDA topics top50 words and query words.}
  \label{fig-overlap}
\end{figure}

To further check this assumption, we conduct comparison with topic model. Without loss of generality, we choose the state-of-the-art, LDA model~\cite{blei2003latent} to generate topic words. LDA model is trained on the whole corpus, with 50 topics and default parameters in package \textsc{gensim}~\cite{rehurek_lrec}. Then we collect top 50 words in each topic, totally 853 words (some words lay in multiple topics). Meanwhile, we collect all pooling words using the Arc-I model, and sort decreasing by the words frequency. The Figure~\ref{fig-overlap} shows the distribution of the word overlap ratio between top pooling words and the top 50 LDA topic words. From the results as we can see, 80\% of top 500 pooling words come from the top 50 LDA topic words, for example in Figure~\ref{fig.arc1-pooling-text}, ``national'' and ``park'' come from the topic 48 and ``office'' come from the topic 36. Additionally, we also measure the overlap ratio between top pooling words and query words (green line in Figure~\ref{fig-overlap}). Contrast with LDA topic words, the overlap ratio of query words is relative small, for example about 30\% of top 500 pooling words come from the query terms.

With above observation, we can conclude that representation-focused models, eg. Arc-I, generate topic related text representation as the topic model do.

\begin{figure}
  \centering
  \includegraphics[width=0.48\textwidth]{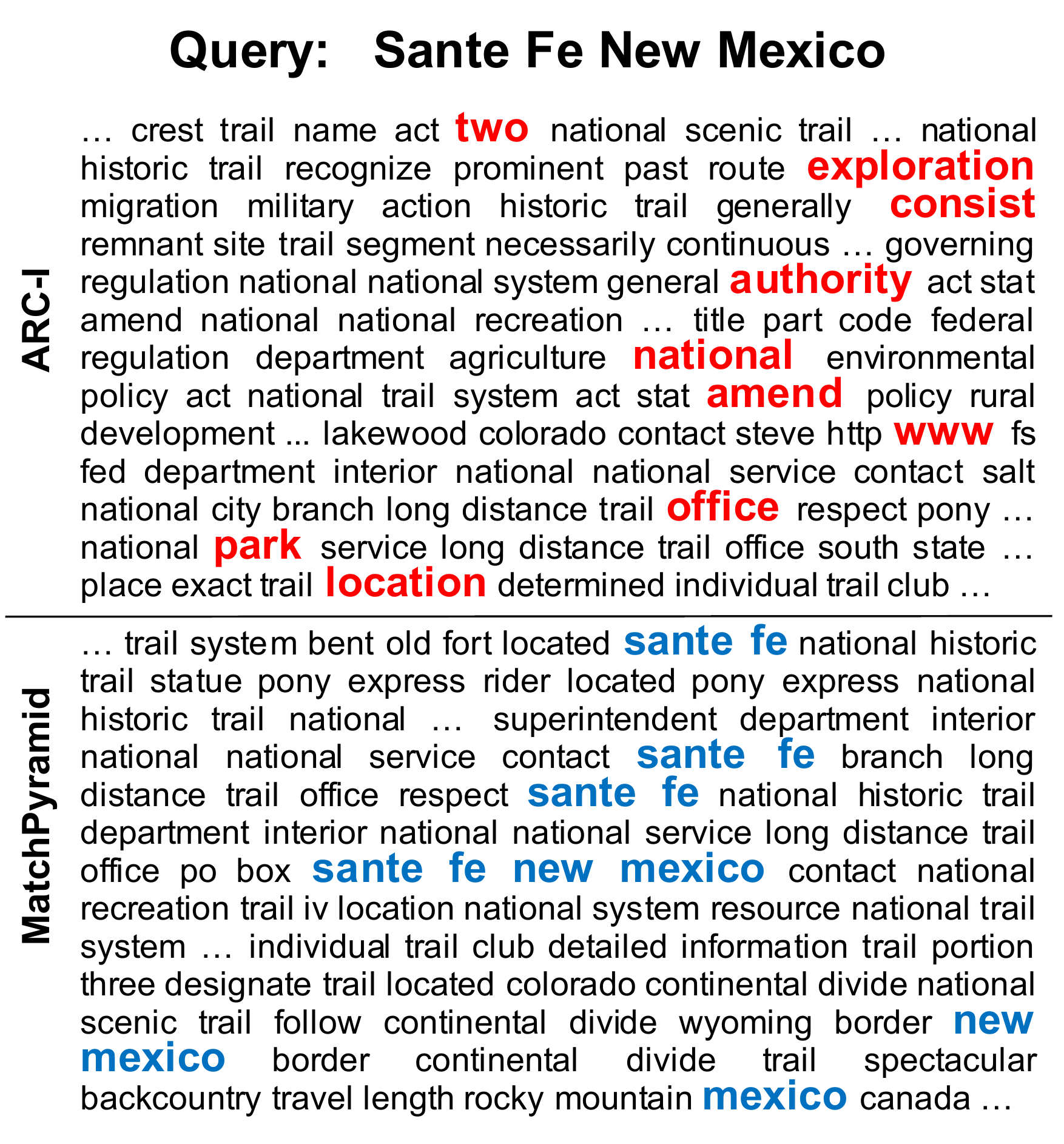}
  \caption{The different pooling words, highlighted using bold font, in the model of Arc-I and MatchPyramid.}
  \label{Fig.pooling_words}
\end{figure}

\subsubsection{Interaction Representations}
Interaction-focused models aim to extract interaction representations on the top of interaction matrix, which is constructed using word-by-word similarity. In paraphrase identification, interaction representation can be interpreted as a hierarchical matching signal composition process, as described in \cite{MatchPyramid}. Firstly, word level matching signals can be composited to phrase level matching signals, for example n-grams matching and n-terms matching. Then, conduct several composition steps to achieve sentence level matching signal. However, it differs a lot when we apply MatchPyramid in information retrieval task. Query is too short to be treated as a sentence, for instance, query ``tooth fairy museum'' only can be treated as a phrase or query ``guatemala'' only has one word. So that hierarchical matching signal composition reduces to word/phrase matching signals aggregation. 

In order to understand what interaction-focused models have learnt, we analyze the MatchPyramid model from the pooling words. Different from the Arc-I model, given a query document pair, almost all pooling words of MatchPyramid come from the query terms. Figure~\ref{Fig.pooling_words} shows the different words pooling by Arc-I and MatchPyramid. 
In MatchPyramid, we use cosine similarity function to evaluate words similarity, thus identical words in query and document achieve the highest similarity value as 1. Then the following convolutional and pooling operations have a high probability to pool out these words.

\subsubsection{Synthetic Analysis}
From the above discussion, we realize that presentation-focused models and interaction-focused models capture different kinds of informations in IR task. In order to distinguish the performance of text representations and interaction representations, we introduce an additional synthetic dataset to verify the ability of two kinds of deep IR models.

The synthetic dataset is constructed using two well-designed ground truth.
\begin{enumerate}
	\item \textbf{Topic Match} The relevance document contain a specific topic, where each topic expresses as a sequential of words. For example, word sequence ``neural network'' represent a topic, and any document contains this word sequence lays into this topic, marked as relevance.
	\item \textbf{Density Match} The relevance degree of a document is proportional to the density of query terms it contained. For example, the document contain 100 query terms is more relevant compare to the one contain 10 query terms.
\end{enumerate}

The whole dataset contains 10,000 queries. In each query, words are randomly sampled from a 2000 words vocabulary. Then, for each query we construct one relevant document and four irrelevant documents. In each document, words are randomly sampled from the same vocabulary with query. The length of query is ranging from 2 to 8, while the length of the document is ranging from 300 to 700.

\begin{figure}
\centering 
\begin{tikzpicture}
 \begin{axis}[
  ybar,
  axis on top,
        height=.30\textwidth,
        width=.45\textwidth,
        bar width=0.5cm,
        ytick={0.0,0.2,0.4,0.6,0.8,1.0},
        ymajorgrids, tick align=inside,
        major grid style={draw=white},
        minor y tick num={1},
        %tickwidth=0pt,
        enlarge y limits={value=.1,upper},
        ymin=0, ymax=1,
        %axis x line*=bottom,
        %axis y line*=left,
        axis lines=left,
        %y axis line style={opacity=0},
        %tickwidth=0pt,
        enlarge x limits=0.25,
        legend style={
            at={(0.5,1.1)},
            font=\small,
            anchor=north,
            draw=none,
            legend columns=-1,
            /tikz/every even column/.append style={column sep=0.1cm}
        },
        ylabel={MAP},
        ylabel style={
            anchor=south,
            at={(ticklabel* cs:1.06)},
            yshift=0pt
        },
        symbolic x coords={ 
        Random,Arc-I,MatchPyramid },
       xtick=data,
       nodes near coords={
        \pgfmathprintnumber[fixed zerofill,precision=2]{\pgfplotspointmeta}
       }
 ]
 \addplot [draw=none, fill=c2, postaction={pattern=north east lines}] coordinates {
      (Random,0.457)
      (Arc-I,0.711)
      (MatchPyramid,0.461)
      };
   \addplot [draw=none, fill=c4, postaction={pattern= crosshatch dots}] coordinates {
      (Random,0.457)
      (Arc-I,0.465)
      (MatchPyramid,0.980)
      };
     
   \legend{Topic Match, Dense Match}
 \end{axis}
\end{tikzpicture}
\caption{Performance comparison of Random, Arc-I and MatchPyramid on synthetic data.}
\label{Figure.Synthetic}
\end{figure}
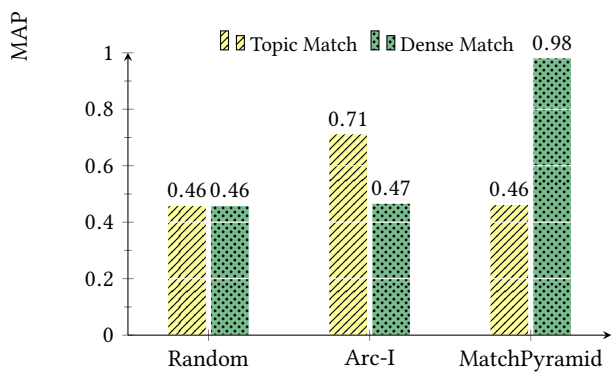

Arc-I and MatchPyramid are evaluated on these two dataset respectively, and the results are shown in Figure~\ref{Figure.Synthetic}. As we can see, the performance of representation-focused model Arc-I and interaction-focused model MatchPyramid is quite the opposite. The representation-focused model have ability to learn topic information, while very little information about query terms density. On the contrast, interaction-focused model is good at learning query terms density information, but almost no topic information covered.

\section{Conclusions and Future Work}\label{sec:conclusion}

As a conclusion, deep IR models still perform worse than hand-crafted features. The possible reason is that under a limit size of dataset,  hand-crafted features, such as BM25, obey the heuristic retrieval constrains, while deep IR models ignored. Apart from using larger dataset, we establish guidelines to explicit using heuristic retrieval constrains, in order to further improvement of deep IR models.

On comparing representation-focused models and interaction-focused models, we can conclude that, 
1) representation-focused models focus on learning good text representation, which encode the topic related words in the final representation; 
2) interaction-focused models focus on learning good interaction representation, which good at collecting the density of matching signals.
These interesting findings pave a way to better understand the different deep IR models and demonstrate that for different applications we need to choose different deep IR models.

%\end{document}  % This is where a 'short' article might terminate

%ACKNOWLEDGMENTS are optional
%\section{Acknowledgments}

\bibliographystyle{acm_ref}
\bibliography{sigproc_short}  % sigproc.bib is the name of the Bibliography in this case
% ACM needs 'a single self-contained file'!
%\balancecolumns
\appendix 
% Appendix A

%\balancecolumns % GM June 2007
% That's all folks!
\end{document}